\pdfoutput=1
\RequirePackage{ifpdf}
\ifpdf % We~are running pdfTeX in pdf mode
\documentclass[pdftex]{sigma}
\else
\documentclass{sigma}
\fi

\usepackage{tikz}
\tikzset{
 font={\fontsize{16pt}{12}\selectfont}}

\usepackage{xcolor}

\numberwithin{equation}{section}

\newtheorem*{Theorem*}{Theorem}

 { \theoremstyle{definition}

 }

\begin{document}
\allowdisplaybreaks

\newcommand{\arXivNumber}{2506.21789}

\renewcommand{\PaperNumber}{077}

\FirstPageHeading

\ShortArticleName{Integrable 3-Site, Extended Bose--Hubbard Model}

\ArticleName{Integrable 3-Site, Tilted, Extended Bose--Hubbard\\ Model with Nearest-Neighbour Interactions}

\Author{Jon LINKS}

\AuthorNameForHeading{J.~Links}

\Address{School of Mathematics and Physics, The University of Queensland, 4072, Australia}
\Email{\href{mailto:jrl@maths.uq.edu.au}{jrl@maths.uq.edu.au}}
\URLaddress{\url{https://people.smp.uq.edu.au/JonLinks/}}

\ArticleDates{Received June 30, 2025, in final form September 12, 2025; Published online September 19, 2025}

\Abstract{Extended Bose--Hubbard models have been employed in the study of cold-atom systems with dipolar interactions. It is shown that, for a certain choice of the coupling parameters, there exists an integrable extended 3-site Bose--Hubbard model with nearest-neighbour interactions. A Bethe ansatz procedure is developed to obtain expressions for the energy spectrum and eigenstates.}


\Classification{17B80; 81R12}

\section{Introduction}

There is consensus that the Bose--Hubbard model on three or more sites with open boundary conditions is not an integrable system, due to the display of chaotic behaviours \cite{bmmbk20,fp03,krbl10,k16,kb04,nh23,ol07}. Extended Bose--Hubbard models, which accommodate quadratic number operator interaction terms between different sites, have received attention for their role in modelling cold-atom systems with dipolar interactions \cite{cblmz25,lmslp09,lps10,rbjg24}. From the mathematical perspective, these extended models also open avenues for constructing integrable generalisations of the Bose--Hubbard model. A broad class of integrable, extended Bose--Hubbard models associated with complete bipartite graphs was formulated in \cite{ytfl17}, generalising the 2-site case \cite{lfts06}. Within this class there is a 3-site model, first studied in \cite{wytlf18} with emphasis on the response to an integrability-breaking tilting of the potential. This model has been studied further to characterise the interface between quantum chaos and integrability \cite{ccrsh21,cwcrfh24,wcfs22,wcrfh25}, and
in relation to entanglement generation~\cite{tywfl20,wyblf23}.

There exists integrable 3-site Bose--Hubbard models with periodic boundary conditions, such as the homogeneous trimer studied in \cite{plwd23} and the non-hermitian system with unidirectional hopping \cite{zqwcc24}. One of the distinctive features of the model of \cite{wytlf18} is that it has open boundary conditions with respect to the tunneling terms (i.e., no tunneling between sites 1 and 3), yet closed boundary conditions with respect to the intersite quadratic number operator interactions (i.e., site 1 couples to site 3).
While it is feasible to engineer a potential to simulate such a Hamiltonian, as discussed in \cite{tywfl20,wytlf18}, it is of complementary interest to consider integrable Hamiltonian open chains where the interactions are restricted to being on-site or nearest-neighbour only. In this work, it will be demonstrated how this can be achieved for a 3-site system. Remarkably, the construction provides for the inclusion of a tilting of the potential that does not break integrability. However, there is a compensation to be paid, viz.\ the model is not homogeneous with respect to the on-site interactions strength. On the other hand, such a property can in principle be accommodated due to the dipolar properties of the constituent particles \cite{cblmz25,lmslp09,lps10,rbjg24}.

The integrable Hamiltonian is introduced in Section \ref{sec2}. It is demonstrated how the Hamiltonian may be formulated via the
$\mathfrak{o}(4)=\mathfrak{o}(3)\oplus \mathfrak{o}(3)$ Lie algebra, realised by canonical boson operators. Through this construction the conserved operators for the Hamiltonian are identified. It is also instructive to view this approach through the lens of the representation theory of $\mathfrak{o}(4)$, in order to set up a Bethe ansatz solution. In Section~\ref{sec3}, an explicit basis is chosen for the Fock space, in a manner that facilitates the derivation of the Bethe ansatz equations. The roots of these equations charaterise both the energy spectrum and the eigenstates of the system. Concluding remarks are offered in Section~\ref{sec4}, including comments regarding the completeness of the Bethe ansatz solution. The appendix contains some technical calculations required for deriving the Bethe ansatz results.

\section{Hamiltonian and symmetries}\label{sec2}

Let $\bigl\{b_j, b_j^\dagger\mid j=1, 2, 3\bigr\}$ denote bosonic annihilation and creation operators satisfying the canonical commutation relations
\[ \big[b_k,b_k^\dagger\big]=\delta_{jk}I, \qquad [b_j, b_k]=\big[b^\dagger_j, b_k^\dagger\big]=0, \]
where $I$ denotes the identity operator. Set $N_j=b^\dagger_j b_j$ and
$N=N_1+N_2+N_3$.
The Hamiltonian for a tilted, 3-site, extended Bose--Hubbard model with nearest-neighbour interactions has the general form
\begin{align}
H={}&U_1 N_1^2+U_2 N_2^2+U_3N_3^2+U_{12}N_1N_2+U_{23}N_2N_3 \nonumber \\
& +\mu_1 N_1 + \mu_3 N_3 + {\mathcal E}_{12}\bigl(b_1^\dagger b_2 + b_2^\dagger b_1\bigr) + {\mathcal E}_{23}\bigl(b_2^\dagger b_3 + b_3^\dagger b_2\bigr). \label{ham}
\end{align}
The Hamiltonian acts on the Fock space ${\mathcal F}$ spanned by the basis vectors
\begin{align*}
|l,m,n\rangle= \bigl(b_1^\dagger\bigr)^l\bigl(b_2^\dagger\bigr)^m \bigl(b_3^\dagger\bigr)^n |0\rangle,
\end{align*}
where $|0\rangle$ denotes the Fock vacuum. For the manipulations below, it is necessary to work with a~particular choice of {\it non-normalised} Fock vectors; this is the reason for the omission of normalisation coefficients.

\begin{figure}[t!]
\centering
\begin{tikzpicture}[scale=0.9]
\draw [cyan, line width=4] plot [smooth, tension=1] coordinates { %(-6.5,2)
(-4.8,1.3) (-2,-4)
 (0,1) (2,-3) (3.5,1)
(6,-2) (8.7,1.3) %(9.5,2)
};
\fill[green] (2,-2.3) circle (0.35);
\draw[black,thick] (2,-2.3) circle (0.35);
\draw[->, line width=1.5] (2,-2.6) -- (2,-2.0);
\fill[green] (1.8,-1.5) circle (0.35);
\draw[black,thick] (1.8,-1.5) circle (0.35);
\draw[->, line width=1.5] (1.8,-1.8) -- (1.8,-1.2);
\fill[green] (2.2,-0.7) circle (0.35);
\draw[black,thick] (2.2,-0.7) circle (0.35);
\draw[->, line width=1.5] (2.2,-1) -- (2.2,-0.4);
%%%%%%%%%%%%%%%%%%%%%%%%%%%%%%%%%%%%%%%%%%%%%%%%%%
\fill[green] (-2.5,-2.7) circle (0.35);
\draw[black,thick] (-2.5,-2.7) circle (0.35);
\draw[->, line width=1.5] (-2.5,-3.0) -- (-2.5,-2.4);
\fill[green] (-1.65,-2.75) circle (0.35);
\draw[black,thick] (-1.65,-2.75) circle (0.35);
\draw[->, line width=1.5] (-1.65,-3.05) -- (-1.65,-2.45);
\fill[green] (-2,-3.5) circle (0.35);
\draw[black,thick] (-2,-3.5) circle (0.35);
\draw[->, line width=1.5] (-2,-3.8) -- (-2,-3.2);
\fill[green] (-2.1,-2.0) circle (0.35);
\draw[black,thick] (-2.1,-2.0) circle (0.35);
\draw[->, line width=1.5] (-2.1,-2.3) -- (-2.1,-1.7);
\fill[green] (-3,-2.1) circle (0.35);
\draw[black,thick] (-3,-2.1) circle (0.35);
\draw[->, line width=1.5] (-3,-2.4) -- (-3,-1.8);
%%%%%%%%%%%%%%%%%%%%%%%%%%%%%%%%%%%%%%%%%%%%%%%%%%%
\fill[green] (6.7,-1.1) circle (0.35);
\draw[black,thick] (6.7,-1.1) circle (0.35);
\draw[->, line width=1.5] (6.7,-1.4) -- (6.7,-0.8);
\fill[green] (6.1,-0.5) circle (0.35);
\draw[black,thick] (6.1,-0.5) circle (0.35);
\draw[->, line width=1.5] (6.1,-0.8) -- (6.1,-0.2);
\fill[green] (6,-1.5) circle (0.35);
\draw[black,thick] (6,-1.5) circle (0.35);
\draw[->, line width=1.5] (6,-1.8) -- (6,-1.2);
\fill[green] (5.4,-0.9) circle (0.35);
\draw[black,thick] (5.4,-0.9) circle (0.35);
\draw[->, line width=1.5] (5.4,-1.2) -- (5.4,-0.6);
\draw[<->, line width=2] (2.8,-1.5) -- (5.0,-1.5);
\draw[<->, line width=2] (-0.9,-2.3) -- (1.3,-2.3);
\draw[<-, line width=2] (2,-4) -- (2,-3.1);
\draw[->, line width=2] (6,-3) -- (6,-2.1);
\draw (2.3, -3.6) node {\small $\mu_1$};
\draw (6.3,-2.6) node {\small $\mu_3$};
\draw (0.2,-2.7) node {\small ${\mathcal E}_{12}$};
\draw (0.2,-1.9) node {\small $U_{12}$};
\draw (3.9,-1.9) node {\small ${\mathcal E}_{23}$};
\draw (3.9,-1.1) node {\small $U_{23}$};
\draw (-2,0.5) node {\small $U_{1}$};
\draw (2,0.5) node {\small $U_{2}$};
\draw (6,0.5) node {\small $U_{3}$};
\end{tikzpicture}
\caption{Schematic representation for a system of dipolar bosons tunneling in a tilted 3-site potential. The narrow profile of the potential at site 2 leads to attractive on-site dipole interaction, such that $U_2$ may be tuned to be a lower value than $U_1$ and $U_3$.}\label{fig}
\end{figure}
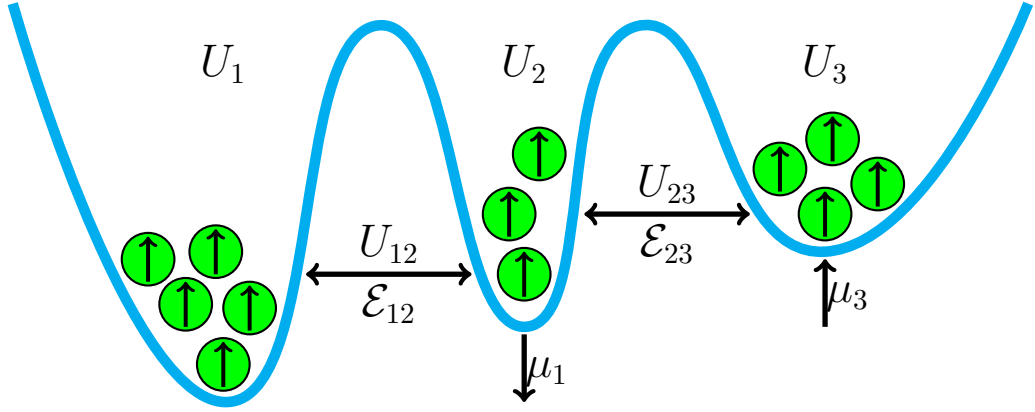

Extended Bose--Hubbard Hamiltonians have been studied in recent times as models for systems comprised of dipolar particles \cite{cblmz25,lmslp09,lps10,rbjg24}. Progress in experimental techniques offers a~remarkable level of control over dipolar systems via the capacity to tune the Hamiltonian coupling parameters through the dipolar interactions. For the 3-site case, a schematic representation is provided in Figure~\ref{fig}. While the figure represents the trapping potential as a~two-dimensional image, it is important to emphasise that in a physical setup each site is accommodated by an ellipsoidal, three-dimensional, localised potential. The on-site dipole interaction, along with on-site contact interactions, are represented by the couplings $U_1$, $U_2$, $U_3$. These can be varied from negative through to positive values, by changing the shape of the ellipsoid from prolate to oblate \cite{klmfgp08}. In addition to influencing interactions on-site, the collective dipole at each site induces interactions between sites, represented by the couplings $U_{12}$, $U_{23}$. It is this level of control in adjusting the interaction parameters that renders dipolar systems as candidates for realising integrable systems, whereby the interaction parameters need to be finely tuned to reach integrable limits of the system.

It is straightforward to verify that \eqref{ham} conserves the total particle number $N$, viz.\ that $[H, N]=0$ holds.
Consequently there is the decomposition
\[
{\mathcal F}=\bigoplus_{n=0}^\infty {\mathcal F}_{\mathcal N},
\]
where ${\mathcal F}_{\mathcal N}={\rm span}\{|l,m,n\rangle \mid l+m+n={\mathcal N}\}$, and each non-zero element of ${\mathcal F}_{\mathcal N}$ is an eigenvector of $N$ with eigenvalue ${\mathcal N}$. The dimensions of the components in the decomposition are
given by the triangular numbers ${\mathcal T}_{\mathcal N}$ through
\begin{align*}
\dim ({\mathcal F}_{\mathcal N})= {\mathcal T}_{{\mathcal N}+1}= \frac{({\mathcal N}+1)({\mathcal N}+2)}{2}.
\end{align*}
Below it will be shown that the Hamiltonian possesses an additional conserved operator under the constraints
\begin{gather}
{\mathcal E}_{12}={\mathcal E}_{23} = {\mathcal E},
\label{int1}  \\
U_1=2U_2=U_3=U_{12}=U_{23}=2U,
\label{int2} \\
\mu_3=-\mu_1=\mu.
\label{int4}
\end{gather}

In order to expose the symmetries underlying this specialisation of \eqref{ham} set
 \begin{alignat*}{4}
& e_1=\sqrt{2}\bigl(b_1^\dagger b_2+b_2^\dagger b_3\bigr),\qquad && h_1= 2(N_1-N_3) ,\qquad&& f_1= \sqrt{2}\bigl(b_2^\dagger b_1+b_3^\dagger b_2\bigr), &\\
&e_2=\frac{1}{2}b_2^\dagger b_2^\dagger -b_1^\dagger b_3^\dagger,\qquad && h_2= \frac{1}{2}(2N+3I) ,\qquad && f_2= b_1b_3-\frac{1}{2}b_2b_2 .&
\end{alignat*}
These operators realise the $\mathfrak{o}(4)$ Lie algebra by satisfying the commutation relations
\begin{align*}
[e_j, f_k]=\delta_{jk}h_j, \qquad
[h_j, e_k]=2\delta_{jk}e_k, \qquad
[h_j, f_k]=-2\delta_{jk}f_k.
\end{align*}
It is found that the Hamiltonian \eqref{ham} subject to the constraints
\eqref{int1}, \eqref{int2}, \eqref{int4} is expressible~as
\begin{align}
H = H_1+H_2,  \qquad
H_1=\frac{U}{4}h_1^2   -\frac{\mu}{2} h_1  +\frac{{\mathcal E}}{\sqrt{2}}(e_1+f_1), \qquad
H_2=U\left(h_2-\frac{3}{2}I\right)^ 2 . \label{hamint}
\end{align}
Under this realisation, the corresponding $\mathfrak{o}(4)$ Casimir invariants assume the form
\begin{gather*}
C_1=\frac{1}{2}h_1^2+e_1f_1+f_1e_1 = 2N^2 +2N -8N_1N_3-2N_2^2 +2N_2 + 4b_2^\dagger b_2^\dagger b_1b_3+4b_1^\dagger b_3^\dagger b_2 b_2  ,\\
C_2=\frac{1}{2} h_2^2+e_2f_2+f_2e_2 = \frac{1}{4}C_1 -\frac{3}{8}I.
\end{gather*}
This realisation of the Casimir invariants provides the third independent conserved operator required to claim integrability of \eqref{hamint}.

\section{Bethe ansatz solution}\label{sec3}

To derive the Bethe ansatz solution of the model, differential operator methods are adapted from \cite{lh06,lz09}.
Before proceeding to that task, it is first necessary to completely describe the~$\mathfrak{o}(4)$ action on the module provided by the Fock space. This exercise belongs to a class of well-known problems in developing a suitable
symmetry-adapted basis with respect to a subalgebra embedding of $\mathfrak{o}(3) \subset \mathfrak{gl}(3)$. See, for example, \cite{mpsw75}.

With respect to the action of $\mathfrak{o}_1(3)={\rm span}\{e_1, h_1, f_1\}$, the Fock space decomposes into a~direct sum of irreducible modules with lowest weight vectors $|0,0,m\rangle$ such that
$
f_1|0,0,n\rangle =0$, $
h_1|0,0,n \rangle = -2n |0,0,n \rangle$, $
C_1|0,0,n \rangle = 2n(n+1) |0,0,n \rangle$.
Such vectors are simultaneously lowest-weight vectors with respect to the action of $\mathfrak{o}_2(3)={\rm span}\{e_2, h_2, f_2\}$, satisfying
\begin{gather*}
f_2|0,0,n\rangle =0, \qquad
h_2|0,0,n\rangle = \frac{1}{2}(2n+3) |0,0,n\rangle, \\
C_2|0,0,n\rangle = \frac{1}{2}\left(n(n+1)-\frac{3}{4}\right) |0,0,n\rangle.
\end{gather*}
Now set, for $({\mathcal N}-n)/2 \in {\mathbb Z}_{\geq 0}$, the recursive definition
\begin{align*}
|n, 0, n\}=|0,0,n\rangle , \qquad
|{\mathcal N}+2, 0, n\} = \frac{-2}{{\mathcal N}+n+3}e_2|{\mathcal N}, 0, n\}.
\end{align*}
The following hold:
\begin{gather*}
f_1 |{\mathcal N}, 0, n\} =0, \qquad
h_1 |{\mathcal N}, 0, n\} = -2n |{\mathcal N}, 0, n\}  ,\qquad
C_1 |{\mathcal N}, 0, n\} = 2n(n+1) |{\mathcal N}, 0, n\}, \\
h_2 |{\mathcal N}, 0, n\} = \frac{1}{2}(2{\mathcal N}+3) |{\mathcal N}, 0, n\}, \qquad
C_2 |{\mathcal N}, 0, n\} = \frac{1}{2}\left(n(n+1)-\frac{3}{4}\right) |{\mathcal N}, 0, n\}.
\end{gather*}
Similarly recursively define, for $m=0,\dots, 2n-1$,
\[
|{\mathcal N}, m+1, n\} = \frac{1}{2n-m}e_1 |{\mathcal N}, m, n\}.
\]
Then the action of the $\mathfrak{o}(4)$ algebra on this set of states is given by (recall $({\mathcal N}-n)/2 \in {\mathbb Z}_{\geq 0}$)
\begin{gather*}
f_1 |{\mathcal N}, m, n\} =m |{\mathcal N}, m-1, n\},\qquad
h_1 |{\mathcal N}, m, n\} = 2(m-n) |{\mathcal N}, m, n\}  , \\
e_1 |{\mathcal N}, m, n\} = (2n-m)|{\mathcal N}, m+1, n\}  , \qquad
C_1 |{\mathcal N}, m, n\} = 2n(n+1) |{\mathcal N}, m, n\}, \\
f_2 |{\mathcal N}, m, n\} =
\frac{{\mathcal N}-n} {2} |{\mathcal N}-2, m, n\} , \qquad
h_2 |{\mathcal N}, m, n\} = \frac{1}{2}(2{\mathcal N}+3) |{\mathcal N}, m, n\}, \\
e_2 |{\mathcal N}, m, n\} = -\frac{{\mathcal N}+n+3}{2} |{\mathcal N}+2, m, n\} ,  \\
C_2 |{\mathcal N}, m, n\} = \frac{1}{2}\left(n(n+1)-\frac{3}{4}\right)
|{\mathcal N}, m, n\}.
\end{gather*}
Since the eigenvalues with respect to $h_1$, $h_2$, $C_1$ uniquely identify $|{\mathcal N}, m, n\}$, it follows that the set of such vectors is linearly independent. That they span ${\mathcal F}$ follows by a counting argument. Each vector $|{\mathcal N},0,n\}$ is a lowest-weight vector of weight $-2n$, so it generates an irreducible $\mathfrak{o}_1(3)$-model $V_n$ of dimension $\dim (V_n)=2n+1$. Now if ${\mathcal N}$ is even, then $n$ is even, and
\begin{align*}
\sum_{n\in 2{\mathbb Z}_{\geq 0}}^{\mathcal N} \dim (V_n)=
\sum_{k=0}^{{\mathcal N}/2} \dim (V_{2k})=\sum_{k=0}^{{\mathcal N}/2} (4k+1)=
 \frac{({\mathcal N}+1)({\mathcal N}+2)}{2}=\dim ({\mathcal F}_{\mathcal N}).
\end{align*}
Otherwise, if ${\mathcal N}$ is odd such that $p=n-1$ is even,
\begin{align*}
\sum_{p\in 2{\mathbb Z}_{\geq 0}}^{{\mathcal N}-1} \dim (V_{p+1})=
\sum_{k=0}^{({\mathcal N}-1) /2} \dim (V_{2k+1})=\sum_{k=0}^{({\mathcal N}-1)/2} (4k+3)=
 \frac{({\mathcal N}+1)({\mathcal N}+2)}{2}=\dim ({\mathcal F}_{\mathcal N}).
\end{align*}
Having established that the vectors $| {\mathcal N}, m, n\} $ provide a basis for ${\mathcal F}$,
define the ${\mathcal N}$-particle states
\begin{gather}
|\Psi(u_1,\dots,u_{2n};{\mathcal N})\rangle=\prod_{j=1}^{2n}({\overline e}_1-u_jI) |{\mathcal N}, 0, n\} ,
\label{states}\\
|\Psi_k(u_1,\dots,u_{2n};{\mathcal N})\rangle=\prod_{j\neq k}^{2n}({\overline e}_1-u_jI) |{\mathcal N}, 0, n\}.
\nonumber
\end{gather}
Above, ${\overline e}_1$ is defined by the action
\begin{align*}
{\overline e}_1 |{\mathcal N},m,n\}= |{\mathcal N},m+1,n\}, \qquad m=0,\dots, 2n-1.
\end{align*}
It is found that
\begin{gather}
e_1|\Psi(u_1,\dots,u_{2n};{\mathcal N})\rangle
=-\sum_{k=1}^{2n} u_k |\Psi(u_1,\dots,u_{2n};{\mathcal N})\rangle
-\sum_{k=1}^{2n} u^2_k |\Psi_k(u_1,\dots,u_{2n};{\mathcal N})\rangle ,
\label{eact}\\
h_1|\Psi(u_1,\dots,u_{2n};{\mathcal N})\rangle
=2n |\Psi(u_1,\dots,u_{2n};{\mathcal N})\rangle +2\sum_{k=1}^{2n} u_k |\Psi_k(u_1,\dots,u_{2n};{\mathcal N})\rangle ,
\label{hact}\\
h_1^2 |\Psi(u_1,\dots,u_{2n};{\mathcal N})\rangle= 4n^2 |\Psi(u_1,\dots,u_{2n};{\mathcal N})\rangle+8\sum_{k=1}^{2n} \sum_{l\neq k}^{2n} \frac{u^2_k}{u_k-u_l} |\Psi_{k}(u_1,\dots,u_{2n};{\mathcal N})\rangle \nonumber \\
\phantom{h_1^2 |\Psi(u_1,\dots,u_{2n};{\mathcal N})\rangle=}{}
-4(2n-1)\sum_{k=1}^{2n} u_k |\Psi_{k}(u_1,\dots,u_{2n};{\mathcal N})\rangle
  , \label{h2act} \\
f_1 |\Psi(u_1,\dots,u_{2n};{\mathcal N})\rangle =\sum_{k=1}^{2n} |\Psi_k(u_1,\dots,u_{2n};{\mathcal N})\rangle .
\label{fact}
\end{gather}
As the calculations leading to the above formulae are somewhat technical, the details have been placed in Appendix \ref{appA}.

The above show that the action of \eqref{hamint} on \eqref{states} is evaluated as
\begin{gather*}
H |\Psi(u_1,\dots,u_{2n};{\mathcal N})\rangle\\
\qquad= U{\mathcal N}^2 |\Psi(u_1,\dots,u_{2n};{\mathcal N})\rangle+
\frac{U}{4}\Biggl(4n^2 |\Psi(u_1,\dots,u_{2n};{\mathcal N})\rangle  \\
  \phantom{\qquad=}{}+ 8\sum_{k=1}^{2n} \sum_{l\neq k}^{2n}  \frac{u^2_k}{u_k-u_l} |\Psi_{k}(u_1,\dots,u_{2n};{\mathcal N})\rangle
-
4(2n-1)\sum_{k=1}^{2n} u_k |\Psi_{k}(u_1,\dots,u_{2n};{\mathcal N})\rangle
	\Biggr) \nonumber \\
 \phantom{\qquad=}{}
 -\frac{\mu}{2} \left(2n |\Psi(u_1,\dots,u_{2n};{\mathcal N})\rangle +2\sum_{k=1}^{2n} u_k |\Psi_k(u_1,\dots,u_{2n};{\mathcal N})\rangle
\right) \\
 \phantom{\qquad=}{}
 -\frac{{\mathcal E}}{\sqrt{2}}\left( \sum_{k=1}^{2n} u_k |\Psi(u_1,\dots,u_{2n};{\mathcal N})\rangle
+\sum_{k=1}^{2n} u^2_k |\Psi_k(u_1,\dots,u_{2n};{\mathcal N})\rangle \right. \\
 \left. \phantom{\qquad=}{} -\sum_{k=1}^{2n} |\Psi_k(u_1,\dots,u_{2n};{\mathcal N})\rangle  \right).
\end{gather*}
The unwanted terms $ |\Psi_{k}(u_1,\dots,u_{2n};{\mathcal N})\rangle $ cancel when the Bethe ansatz equations
\begin{align}
2U\sum_{l\neq k}^{2n}  \frac{u^2_k}{u_k-u_l} &= (U(2n-1)+\mu)u_k+\frac{\mathcal E}{\sqrt{2}} \bigl(u_k^2-1\bigr),
\qquad k=1,\dots,2n,
\label{bae}
\end{align}
hold, rendering $|\Psi(u_1,\dots,u_{2n};{\mathcal N}\rangle$ an eigenstate with energy eigenvalue
\begin{align}
E= U\bigl({\mathcal N}^2+n^2\bigr)-\mu n -\frac{\mathcal E}{\sqrt{2}} \sum_{k=1}^{2n} u_k.
\label{nrg}
\end{align}
Note that the above solution includes the cases $n=0$, for which the state $|{\mathcal N}, 0, 0\}$ is an eigenstate of the Hamiltonian \eqref{ham} with eigenvalue $U{\mathcal N}^2$ for each
${\mathcal N}\in 2{\mathbb Z}_{\geq 0}$.

\section{Conclusion}\label{sec4}

This work reports the construction of an integrable 3-site, extended, Bose--Hubbard model, providing a counterpoint to the integrable model studied in
\cite{ccrsh21,cwcrfh24,tywfl20,wcfs22,wcrfh25,wyblf23,wytlf18}.
Two distinguishing features of the Hamiltonian
\eqref{hamint} are that all interactions are on-site or nearest neighbour, and that the Hamiltonian remains integrable in the presence of a tilting potential with coupling $\mu$. A by-product of the integrability is that the model admits a Bethe ansatz solution.

The Bethe ansatz solution presented above through equations \eqref{states}, \eqref{bae}, and \eqref{nrg} is complete. All eigenstates of the system can be cast into the form \eqref{states}. See Appendix \ref{appB} for details.
Moreover, spurious solutions of \eqref{bae} cannot occur. Spurious solutions of Bethe ansatz equations arise when the evaluation of the Bethe state through the Bethe roots yields a null state. While this is a feature of some spin \cite{a92,av86,gd14,ks14,nw14}, fermionic \cite{lmz13}, and anyonic \cite{cdil10} systems, spurious solutions do not arise here. Since boson creation operators do not admit a non-trivial kernel, the general form
\eqref{states} cannot vanish for any choice of $\{u_1,\dots,u_{2n}\}$.

For future work, one avenue is to investigate the quantum dynamics of the system utilising the Bethe ansatz solution and undertaking a comparison with the results of \cite{ccrsh21,cwcrfh24,tywfl20,wcfs22,wcrfh25,wyblf23,wytlf18}. A particular focus will be to characterise the dynamics in the so-called
resonant tunnelling regime, where there are oscillations that are approximately harmonic. It would be of interest to derive analytic formulae for the amplitude and frequency of these oscillations, to understand their dependency on the tilting parameter $\mu$.

Finally, an important consequence of the formulation above is that it facilitates extension to systems with more degrees of freedom. In particular, a 4-site system is accommodated through a realisation of $\mathfrak{o}(3)\oplus \mathfrak{o}(3) \oplus \mathfrak{o}(3)$, providing a counterpoint to the integrable model studied in~\cite{bil23,gwylf22,gywtfl22} for interferometric applications. A study of this 4-site system, obtained by extending the methods developed above, will be communicated in a forthcoming publication.

\appendix

\section[Action of o\_1(3) on the Bethe states]{Action of $\boldsymbol{\mathfrak{o}_1(3)}$ on the Bethe states}\label{appA}
To derive the formulae \eqref{eact}, \eqref{hact}, \eqref{h2act}, \eqref{fact}, it is useful to exploit a correspondence between the algebraic action and differential operators. Making the identification
$|{\mathcal N},m,n\}\mapsto x^m$, it is seen that
$
	e_1\mapsto 2n x -x^2 \frac{\rm d}{{\rm d}x}$, $
	h_1\mapsto  2x \frac{\rm d}{{\rm d}x} -2n$, $
	f_1\mapsto \frac{\rm d}{{\rm d}x}
$
is a Lie algebra isomorphism, and
${\overline e}_1 \mapsto x
$. It follows that
\begin{align*}
e_1|\Psi(u_1,\dots,u_{2n};{\mathcal N})\rangle &= 2n {\overline e}_1 |\Psi(u_1,\dots,u_{2n};{\mathcal N})\rangle -\sum_{k=1}^{2n} {\overline e}^2_1|\Psi_k(u_1,\dots,u_{2n};{\mathcal N})\rangle \\
&=-\sum_{k=1}^{2n} u_k {\overline e}_1|\Psi_k(u_1,\dots,u_{2n};{\mathcal N})\rangle \\
&=-\sum_{k=1}^{2n} u_k |\Psi(u_1,\dots,u_{2n};{\mathcal N})\rangle
-\sum_{k=1}^{2n} u^2_k |\Psi_k(u_1,\dots,u_{2n};{\mathcal N})\rangle,
\end{align*}
which is equation~\eqref{eact}. Similarly,
\begin{align*}
h_1|\Psi(u_1,\dots,u_{2n};{\mathcal N})\rangle&=2\sum_{k=1}^{2n} {\overline e}_1 |\Psi_k(u_1,\dots,u_{2n};{\mathcal N})\rangle
-2n |\Psi(u_1,\dots,u_{2n};{\mathcal N})\rangle \\
&=2\sum_{k=1}^{2n} u_k |\Psi_k(u_1,\dots,u_{2n};{\mathcal N})\rangle
+2n |\Psi(u_1,\dots,u_{2n};{\mathcal N})\rangle,
\end{align*}
providing \eqref{hact},
while \eqref{fact} follows directly.

The derivation of \eqref{h2act} is more involved. Defining
$
|\Psi_{kl}(u_1,\dots,u_{2n};{\mathcal N})\rangle=\prod_{j\neq k,l}^{2n}({\overline e}_1-u_jI) |{\mathcal N}, 0, n\}$,
$ k\neq l
$
and using
\smash{$
h_1^2\mapsto 4x^2 \frac{{\rm d}^2}{{\rm d}x^2} -4x(2n-1) \frac{\rm d}{{\rm d}x}  +4n^2$},
results in
 \begin{gather*}
h^2_1 |\Psi(u_1,\dots,u_{2n};{\mathcal N})\rangle \\
\qquad=4\sum_{k=1}^{2n} \sum_{l\neq k}^{2n} {\overline e}^2_1 |\Psi_{kl}(u_1,\dots,u_{2n};{\mathcal N})\rangle
-4(2n-1)\sum_{k=1}^{2n} {\overline e}_1 |\Psi_k(u_1,\dots,u_{2n};{\mathcal N})\rangle \\
\phantom{\qquad=}{} +4n^2 |\Psi(u_1,\dots,u_{2n};{\mathcal N})\rangle \\
\qquad=4\sum_{k=1}^{2n} \sum_{l\neq k}^{2n} {\overline e}^2_1 |\Psi_{kl}(u_1,\dots,u_{2n};{\mathcal N})\rangle
- 4\sum_{k=1}^{2n} \sum_{l\neq k}^{2n} {\overline e}_1({\overline e}_1-u_l) |\Psi_{kl}(u_1,\dots,u_{2n};{\mathcal N})\rangle \\
\phantom{\qquad=}{} +4n^2 |\Psi(u_1,\dots,u_{2n};{\mathcal N})\rangle \\
\qquad=4\sum_{k=1}^{2n} \sum_{l\neq k}^{2n} u_l {\overline e}_1 |\Psi_{kl}(u_1,\dots,u_{2n};{\mathcal N})\rangle
 +4n^2 |\Psi(u_1,\dots,u_{2n};{\mathcal N})\rangle .
\end{gather*}
Elimination of the $|\Psi_{kl}(u_1,\dots,u_{2n};{\mathcal N})\rangle$ terms is achieved through
 \begin{gather*}
\sum_{k=1}^{2n} \sum_{l\neq k}^{2n}  u_l {\overline e}_1 |\Psi_{kl}(u_1,\dots,u_{2n};{\mathcal N})\rangle \\
\qquad=\sum_{k=1}^{2n} \sum_{l\neq k}^{2n} (u_k+u_l) {\overline e}_1 |\Psi_{kl}(u_1,\dots,u_{2n};{\mathcal N})\rangle
-\sum_{k=1}^{2n} \sum_{l\neq k}^{2n} u_k u_l |\Psi_{kl}(u_1,\dots,u_{2n};{\mathcal N})\rangle \\
\phantom{\qquad=}{} -\sum_{k=1}^{2n} \sum_{l\neq k}^{2n} u_k({\overline e}_1-u_l) |\Psi_{kl}(u_1,\dots,u_{2n};{\mathcal N})\rangle \\
\qquad=\sum_{k=1}^{2n} \sum_{l\neq k}^{2n} \frac{u^2_k-u^2_l}{u_k-u_l} {\overline e}_1|\Psi_{kl}(u_1,\dots,u_{2n};{\mathcal N})\rangle
-\sum_{k=1}^{2n} \sum_{l\neq k}^{2n} \frac{u^2_k u_l- u_ku_l^2}{u_k-u_l} |\Psi_{kl}(u_1,\dots,u_{2n};{\mathcal N})\rangle \\
\phantom{\qquad=}{}-(2n-1)\sum_{k=1}^{2n} u_k |\Psi_{k}(u_1,\dots,u_{2n};{\mathcal N})\rangle \\
\qquad=\sum_{k=1}^{2n} \sum_{l\neq k}^{2n}  \frac{u^2_k}{u_k-u_l} |\Psi_{k}(u_1,\dots,u_{2n};{\mathcal N})\rangle
-\sum_{l=1}^{2n} \sum_{k\neq l}^{2n}  \frac{u_l^2}{u_k-u_l} |\Psi_{l}(u_1,\dots,u_{2n};{\mathcal N})\rangle \\
\phantom{\qquad=}{} -(2n-1)\sum_{k=1}^{2n} u_k |\Psi_{k}(u_1,\dots,u_{2n};{\mathcal N})\rangle \\
\qquad=2\sum_{k=1}^{2n} \sum_{l\neq k}^{2n}  \frac{u^2_k}{u_k-u_l} |\Psi_{k}(u_1,\dots,u_{2n};{\mathcal N})\rangle
-(2n-1)\sum_{k=1}^{2n} u_k |\Psi_{k}(u_1,\dots,u_{2n};{\mathcal N})\rangle ,
\end{gather*}
leading to \eqref{h2act}.

\section{Proof of completeness of the Bethe ansatz solution}\label{appB}

Here it will be shown that the Bethe ansatz solution is complete, by utilising established methods in the analysis of ordinary differential equations. See, for example, \cite{s05}.

The set of states
\begin{align}
\bigl\{ e_1^j|{\mathcal N},0,n\}\mid j=0,\dots, 2n \bigr\}
\label{sub}
\end{align}
provides a basis of weight states for an $\mathfrak{o}_1(3)$ module of highest weight $2n$. Note the actions
\begin{gather*}
e_1^{2n+1}|{\mathcal N},0,n\}=0, \qquad
h_1e_1^j|{\mathcal N},0,n\}= 2(j-n) e_1^j|{\mathcal N},0,n\}, \\
f_1 e_1^j|{\mathcal N},0,n\} = j(2n+1-j)e_1^{j-1}|{\mathcal N},0,n\}.
\end{gather*}
Since the module spanned by \eqref{sub} is invariant under the action of the Hamiltonian \eqref{hamint}, it follows that all eigenstates of \eqref{hamint} within this subspace may be expressed in the form
\begin{align}
 |\psi\rangle=\left(\sum_{j=0}^{2n} \alpha_{j} e^{2n-j}_1\right) |{\mathcal N},0,n\}, \qquad \alpha_j\in {\mathbb C}.
\label{sumstate}
\end{align}
Now
\begin{align*}
H|\psi\rangle={}&U{\mathcal N}^2\sum_{j=0}^{2n} \alpha_j e^{2n-j}_1 |{\mathcal N},0,n\}+\frac{U}{4}\sum_{j=0}^{2n}
 4(n-j)^2\alpha_j e^{2n-j}_1 |{\mathcal N},0,n\} \\
& -\frac{\mu}{2}\sum_{j=0}^{2n} 2(n-j) \alpha_j e^{2n-j}_1 |{\mathcal N},0,n\} \\
& +\frac{{\mathcal E}}{\sqrt{2}}\sum_{j=1}^{2n} \alpha_j e^{2n-j+1}_1 |{\mathcal N},0,n\}
+\frac{{\mathcal E}}{\sqrt{2}}\sum_{j=0}^{2n-1}  (2n-j)(j+1) \alpha_j e^{2n-j-1}_1 |{\mathcal N},0,n\}.
\end{align*}
In order to satisfy the eigenvalue equation $H|\psi\rangle =E |\psi\rangle$, the following system of recursion equations must be satisfied:
\begin{gather}
\alpha_1 = \frac{\sqrt{2}}{{\mathcal E}} \bigl(E- U\bigl({\mathcal N}^2+n^2\bigr)
+{\mu}n \bigr) \alpha_0,
 \label{hom1}\\
 \alpha_{j+1} = \frac{\sqrt{2}}{{\mathcal E}}\bigl(E - U\bigl({\mathcal N}^2+(n-j)^2\bigr)
+{\mu}(n-j) \bigr) \alpha_j
- j(2n-j+1)\alpha_{j-1}
, \nonumber\\
 \hphantom{\alpha_{j+1} =}{}  j=1,\dots, 2n-1,
\label{hom2}\\
 0 =  \bigl( E  -U\bigl({\mathcal N}^2+n^2\bigr)
-{\mu} n \bigr) \alpha_{2n}-\sqrt{2}{\mathcal E}n\alpha_{2n-1}.
\label{hom3}
\end{gather}
Observe that setting $\alpha_{0}=0$ enforces $\alpha_j=0$ for all $j=1,\dots, 2n$. Without loss of generality, since the system \eqref{hom1}, \eqref{hom2}, \eqref{hom3} is homogeneous, set $\alpha_{0}=1$. Then it is seen from \eqref{hom1}, \eqref{hom2} that each $\alpha_j$ is a polynomial in $E$ of degree $j$. The right-hand
side of equation \eqref{hom3} is thus a polynomial of degree $2n+1$, while the $2n+1$ roots of this polynomial that solve \eqref{hom3} provide the complete spectrum on the subspace spanned by \eqref{sub}.
For each $E$ contained in this complete spectrum, the equations \eqref{hom1}, \eqref{hom2} uniquely determine, subject to
$\alpha_0=1$, the $\alpha_j$ appearing in the corresponding eigenstate \eqref{sumstate}. Thus, the spectrum is simple.

The next step is to show that there is a one-to-one correspondence between the eigenstates of the form
\eqref{sumstate} as described above and the solutions of the Bethe ansatz equations \eqref{bae}.
Set
\begin{align}
Q(x)&=\sum_{j=0}^{2n} \frac{\alpha_j}{j!} x^{2n-j}.
\label{sumpoly}
\end{align}
Recalling that $\alpha_0=1$, $Q(x)$ admits a unique factorisation
\begin{align}
Q(x)&=\prod_{j=1}^{2n}(x-v_j). \label{poly}
\end{align}
As a result of the system of equations \eqref{hom1}, \eqref{hom2}, \eqref{hom3}, it follows that \eqref{sumpoly} satisfies the ordinary differential equation
\begin{gather}
Ux^2Q''(x)+\left((U(1-2n)-\mu) x+\frac{{\mathcal E}}{\sqrt{2}}\bigl(1-x^2\bigr)\right)Q'(x) \nonumber\\
\qquad{}  +\bigl(U\bigl({\mathcal N}^2+n^2\bigr)+\mu n+\sqrt{2}{\mathcal E}nx\bigr)Q(x)=E Q(x).
\label{ode} \end{gather}
Now it is asserted that, for any given eigenvalue $E$, the corresponding roots $\{v_1,\dots,v_{2n}\}$ appearing in \eqref{poly} are distinct. The proof is by contradiction. Supposing that $v_k$ has multiplicity~${m_k>1}$, then
\begin{align*}
\left.\frac{{\rm d}^{p} Q(x)}{{\rm d} x^{p}}\right|_{x=v_k} =
\begin{cases}
0, & p< m_k, \\
\text{non-zero} ,&p=m_k.
\end{cases}
\end{align*}
Differentiating \eqref{ode} $m_k-2$ times and making the substitution $x=v_k$ yields
\begin{align*}
U v_k^2\left.\frac{{\rm d}^{m_k} Q(x)}{{\rm d} x^{m_k}}\right|_{x=v_k} =0,
\end{align*}
imposing that $v_k=0$ for $U\neq 0$. (For $U=0$ the system is diagonalisable by a canonical transformation.) However, if $v_k=0$ is a root of $Q(x)$ then $\alpha_{2n}=0$. The recursion relations~\eqref{hom1}, \eqref{hom2}, \eqref{hom3} subsequently establish that $Q(x)=0$, contradicting the assumption $\alpha_0=1$. Hence for each eigenvalue $E$ the roots of the associated polynomial $Q(x)$ are free of multiplicities.

Finally, setting $u=v_k$ in \eqref{ode} yields
\begin{align}
v_k^2\frac{Q''(v_k)}{Q'(v_k)}&=(4U(2n-1))+\mu) v_k+\frac{{\mathcal E}}{\sqrt{2}}\bigl(v_k^2-1\bigr),
\qquad k=1,\dots,2n.
\label{bae_alt}
\end{align}
Using
\begin{align*}
\frac{Q''(v_k)}{Q'(v_k)}=\sum_{l\neq k}^{2n}\frac{2}{v_k-v_l}
\end{align*}
shows that \eqref{bae_alt} is identical to \eqref{bae}. Moreover, equating the coefficients of the
terms of order~$2n$ in \eqref{ode} produces \eqref{nrg}.

Hence the Bethe ansatz is complete. Each eigenstate \eqref{sumstate} uniquely determines a polynomial~\eqref{sumpoly}, or equivalently \eqref{poly}, whose roots satisfy \eqref{bae}. The correspondence is one-to-one; each solution of \eqref{bae} uniquely determines a polynomial which, expressed in the form
\eqref{sumpoly}, provides a solution
$\{\alpha_0,\dots,\alpha_{2n}\}$ for the recursion relations \eqref{hom1}, \eqref{hom2}, \eqref{hom3}. Since the spectrum of the Hamiltonian is simple, all eigenstates arise in this manner.

\subsection*{Acknowledgments}

This research was supported by the Australian Research Council through Discovery Project DP200101339, {\it Quantum control designed from broken integrability}.
The author thanks Lachlan Bennett, Phil Isaac, Angela Foerster, Sam Kault, Rodrigo Pimenta, Mariana Kehl Scola, and Owen Thompson for discussions, and the anonymous referees for their feedback leading to improvements in the manuscript.
The author acknowledges the traditional owners of the Turrbal and Jagera country on which The University of Queensland (St.\ Lucia campus) operates.

\pdfbookmark[1]{References}{ref}
\LastPageEnding


\begin{thebibliography}{99}
\footnotesize\itemsep=0pt

\bibitem{a92}
Albertini G., Bethe-ansatz type equations for the {F}ateev--{Z}amolodchikov
  spin model,
  \href{https://doi.org/10.1088/0305-4470/25/7/021}{\textit{J.~Phys.~A}}
  \textbf{25} (1992), 1799--1813.

\bibitem{av86}
Avdeev L.V., Vladimirov A.A., Exceptional solutions to the {B}ethe ansatz
  equations, \href{https://doi.org/10.1007/BF01037864}{\textit{Theoret. and
  Math. Phys.}} \textbf{69} (1986), 1071--1079.

\bibitem{bil23}
Bennett L., Isaac P.S., Links J., N{OON} state measurement probabilities and
  outcome fidelities: a {B}ethe ansatz approach,
  \href{https://doi.org/10.1088/1751-8121/ad0a71}{\textit{J.~Phys.~A}}
  \textbf{56} (2023), 505202, 27~pages.

\bibitem{bmmbk20}
Bychek A.A., Muraev P.S., Maksimov D.N., Kolovsky A.R., Bulgakov E.N., Chaotic
  and regular dynamics in the three-site {B}ose--{H}ubbard model,
  \href{https://doi.org/10.1063/5.0011540}{\textit{AIP Conf. Proc.}}
  \textbf{2241} (2020), 020007, 7~pages,
  \href{http://arxiv.org/abs/1910.12489}{arXiv:1910.12489}.

\bibitem{cdil10}
Campbell C.W., Dancer K.A., Isaac P.S., Links J., Bethe ansatz solution of an
  integrable, non-abelian anyon chain with~{$D(D_3)$} symmetry,
  \href{https://doi.org/10.1016/j.nuclphysb.2010.04.014}{\textit{Nuclear
  Phys.~B}} \textbf{836} (2010), 171--185,
  \href{http://arxiv.org/abs/1003.3514}{arXiv:1003.3514}.

\bibitem{ccrsh21}
Castro E.R., Ch\'avez-Carlos J., Roditi I., Santos L.F., Hirsch J.G.,
  Quantum-classical correspondence of a~system of interacting bosons in a
  triple-well potential,
  \href{https://doi.org/10.22331/q-2021-10-19-563}{\textit{Quantum}} \textbf{5}
  (2021), 563--574, \href{http://arxiv.org/abs/2105.10515}{arXiv:2105.10515}.

\bibitem{cwcrfh24}
Castro E.R., Wittmann~W. K., Ch\'avez-Carlos J., Roditi I., Foerster A., Hirsch
  J.G., Quantum-classical correspondence in a triple-well bosonic model: from
  integrability to chaos,
  \href{https://doi.org/10.1103/physreva.109.032225}{\textit{Phys. Rev.~A}}
  \textbf{109} (2024), 032225, 13~pages,
  \href{http://arxiv.org/abs/2311.13189}{arXiv:2311.13189}.

\bibitem{cblmz25}
Chanda T., Barbiero L., Lewenstein M., Mark M.J., Zakrzewski J., Recent
  progress on quantum simulations of non-standard {B}ose--{H}ubbard models,
  \href{https://doi.org/10.1088/1361-6633/adc3a7}{\textit{Rep. Progr. Phys.}}
  \textbf{88} (2025), 044501, 27~pages,
  \href{http://arxiv.org/abs/2405.07775}{arXiv:2405.07775}.

\bibitem{fp03}
Franzosi R., Penna V., Chaotic behavior, collective modes, and self-trapping in
  the dynamics of three coupled {B}ose--{E}instein condensates,
  \href{https://doi.org/10.1103/PhysRevE.67.046227}{\textit{Phys. Rev.~E}}
  \textbf{67} (2003), 046227, 37~pages,
  \href{http://arxiv.org/abs/cond-mat/0203509}{arXiv:cond-mat/0203509}.

\bibitem{gd14}
Giri P.R., Deguchi T., Singular eigenstates in the even (odd) length
  {H}eisenberg spin chain,
  \href{https://doi.org/10.1088/1751-8113/48/17/175207}{\textit{J.~Phys.~A}}
  \textbf{48} (2015), 175207, 26~pages,
  \href{http://arxiv.org/abs/1411.5839}{arXiv:1411.5839}.

\bibitem{gwylf22}
Gr\"un D.S., Wittmann~W. K., Ymai L.H., Links J., Foerster A., Protocol designs
  for {NOON} states,
  \href{https://doi.org/10.1038/s42005-022-00812-7}{\textit{Commun. Phys.}}
  \textbf{5} (2022), 36, 13~pages,
  \href{http://arxiv.org/abs/2102.02944}{arXiv:2102.02944}.

\bibitem{gywtfl22}
Gr\"un D.S., Wittmann~W. K., Ymai L.H., Links J., Foerster A., Tonel A.P.,
  Integrable atomtronic interferometry,
  \href{https://doi.org/10.1103/PhysRevLett.129.020401}{\textit{Phys. Rev.
  Lett.}} \textbf{129} (2022), 020401, 9~pages,
  \href{http://arxiv.org/abs/2004.11987}{arXiv:2004.11987}.

\bibitem{ks14}
Kirillov A.N., Sakamoto R., Singular solutions to the {B}ethe ansatz equations
  and rigged configurations,
  \href{https://doi.org/10.1088/1751-8113/47/20/205207}{\textit{J.~Phys.~A}}
  \textbf{47} (2014), 205207, 20~pages,
  \href{http://arxiv.org/abs/1402.0651}{arXiv:1402.0651}.

\bibitem{klmfgp08}
Koch T., Lahaye T., Metz J., Fr\"ohlich B., Griesmaier A., Pfau T.,
  Stabilization of a purely dipolar quantum gas against collapse,
  \href{https://doi.org/10.1038/nphys887}{\textit{Nature Phys.}} \textbf{4}
  (2008), 218--222, \href{http://arxiv.org/abs/0710.3643}{arXiv:0710.3643}.

\bibitem{krbl10}
Kollath C., Roux G., Biroli G., L\"auchli A.M., Statistical properties of the
  spectrum of the extended {B}ose--{H}ubbard model,
  \href{https://doi.org/10.1088/1742-5468/2010/08/p08011}{\textit{J.~Stat.
  Mech. Theory Exp.}} \textbf{2010} (2010), P08011, 18~pages,
  \href{http://arxiv.org/abs/1004.2203}{arXiv:1004.2203}.

\bibitem{k16}
Kolovsky A.R., Bose--{H}ubbard {H}amiltonian: quantum chaos approach,
  \href{https://doi.org/10.1142/S0217979216300097}{\textit{Internat.~J. Modern
  Phys.~B}} \textbf{30} (2016), 1630009, 21~pages,
  \href{http://arxiv.org/abs/1507.03413}{arXiv:1507.03413}.

\bibitem{kb04}
Kolovsky A.R., Buchleitner A., Quantum chaos in the {B}ose--{H}ubbard model,
  \href{https://doi.org/10.1209/epl/i2004-10265-7}{\textit{Europhys. Lett.}}
  \textbf{68} (2004), 632--638,
  \href{http://arxiv.org/abs/cond-mat/0403213}{arXiv:cond-mat/0403213}.

\bibitem{lmslp09}
Lahaye T., Menotti C., Santos L., Lewenstein M., Pfau T., The physics of
  dipolar bosonic quantum gases,
  \href{https://doi.org/10.1088/0034-4885/72/12/126401}{\textit{Rep. Progr.
  Phys.}} \textbf{72} (2009), 126401, 71~pages,
  \href{http://arxiv.org/abs/0905.0386}{arXiv:0905.0386}.

\bibitem{lps10}
Lahaye T., Santos L., Pfau T., Mesoscopic ensembles of polar bosons in
  triple-well potentials,
  \href{https://doi.org/10.1103/PhysRevLett.104.170404}{\textit{Phys. Rev.
  Lett.}} \textbf{104} (2010), 170404, 8~pages,
  \href{http://arxiv.org/abs/0911.5288}{arXiv:0911.5288}.

\bibitem{lfts06}
Links J., Foerster A., Tonel A.P., Santos G., The two-site {B}ose--{H}ubbard
  model, \href{https://doi.org/10.1007/s00023-006-0295-3}{\textit{Ann. Henri
  Poincar\'e}} \textbf{7} (2006), 1591--1600,
  \href{http://arxiv.org/abs/cond-mat/0605486}{arXiv:cond-mat/0605486}.

\bibitem{lh06}
Links J., Hibberd K.E., Bethe ansatz solutions of the {B}ose--{H}ubbard dimer,
  \href{https://doi.org/10.3842/SIGMA.2006.095}{\textit{SIGMA}} \textbf{2}
  (2006), 095, 8~pages,
  \href{http://arxiv.org/abs/nlin.SI/0612063}{arXiv:nlin.SI/0612063}.

\bibitem{lmz13}
Links J., Moghaddam A., Zhang Y.Z., B{CS} model with asymmetric pair
  scattering: a non-{H}ermitian, exactly solvable {H}amiltonian exhibiting
  generalized exclusion statistics,
  \href{https://doi.org/10.1088/1751-8113/46/30/305205}{\textit{J.~Phys.~A}}
  \textbf{46} (2013), 305205, 18~pages,
  \href{http://arxiv.org/abs/1304.5818}{arXiv:1304.5818}.

\bibitem{lz09}
Links J., Zhao S.Y., A {B}ethe ansatz study of the ground state energy for the
  repulsive {B}ose--{H}ubbard dimer,
  \href{https://doi.org/10.1088/1742-5468/2009/03/p03013}{\textit{J.~Stat.
  Mech. Theory Exp.}} \textbf{2009} (2009), P03013, 16~pages.

\bibitem{mpsw75}
Moshinsky M., Patera J., Sharp R.T., Winternitz P., Everything you always
  wanted to know about~{${\rm SU}(3)\supset {\rm O}(3)$},
  \href{https://doi.org/10.1016/0003-4916(75)90048-2}{\textit{Ann. Physics}}
  \textbf{95} (1975), 139--169.

\bibitem{nh23}
Nakerst G., Haque M., Chaos in the three-site {B}ose--{H}ubbard model:
  classical versus quantum,
  \href{https://doi.org/10.1103/physreve.107.024210}{\textit{Phys. Rev.~E}}
  \textbf{107} (2023), 024210, 14~pages,
  \href{http://arxiv.org/abs/2203.09953}{arXiv:2203.09953}.

\bibitem{nw14}
Nepomechie R.I., Wang C., Twisting singular solutions of {B}ethe's equations,
  \href{https://doi.org/10.1088/1751-8113/47/50/505004}{\textit{J.~Phys.~A}}
  \textbf{47} (2014), 505004, 9~pages,
  \href{http://arxiv.org/abs/1409.7382}{arXiv:1409.7382}.

\bibitem{ol07}
Oelkers N., Links J., Ground-state properties of the attractive one-dimensional
  {B}ose--{H}ubbard models,
  \href{https://doi.org/10.1103/PhysRevB.75.115119}{\textit{Phys. Rev.~B}}
  \textbf{75} (2007), 115119, 17~pages,
  \href{http://arxiv.org/abs/cond-mat/0611510}{arXiv:cond-mat/0611510}.

\bibitem{plwd23}
Pan F., Li A., Wu Y., Draayer J.P., An exact solution of the homogenous trimer
  {B}ose--{H}ubbard model,
  \href{https://doi.org/10.1088/1742-5468/acb7ec}{\textit{J.~Stat. Mech. Theory
  Exp.}} \textbf{2023} (2023), 033101, 21~pages.

\bibitem{rbjg24}
Rovirola M., Briongos-Merino H., Juli\'a-D\'iaz B., Guilleumas M., Ultracold
  dipolar bosons trapped in atomtronic circuits,
  \href{https://doi.org/10.1103/PhysRevA.109.063331}{\textit{Phys. Rev.~A}}
  \textbf{109} (2024), 063331, 8~pages,
  \href{http://arxiv.org/abs/2403.11620}{arXiv:2403.11620}.

\bibitem{s05}
Simon B., Sturm oscillation and comparison theorems, in Sturm-{L}iouville
  theory, \href{https://doi.org/10.1007/3-7643-7359-8_2}{Birkh\"auser}, Basel,
  2005, 29--43,
  \href{http://arxiv.org/abs/math.SP/0311049}{arXiv:math.SP/0311049}.

\bibitem{tywfl20}
Tonel A.P., Ymai L.H., Wittmann~W. K., Foerster A., Links J., Entangled states
  of dipolar bosons generated in a triple-well potential,
  \href{https://doi.org/10.21468/SciPostPhysCore.2.1.003}{\textit{SciPost
  Phys.}} \textbf{2} (2020), 003, 20~pages,
  \href{http://arxiv.org/abs/1909.04815}{arXiv:1909.04815}.

\bibitem{wcfs22}
Wittmann~W. K., Castro E.R., Foerster A., Santos L.F., Interacting bosons in a
  triple well: preface of many-body quantum chaos,
  \href{https://doi.org/10.1103/physreve.105.034204}{\textit{Phys. Rev.~E}}
  \textbf{105} (2022), 034204, 15~pages,
  \href{http://arxiv.org/abs/2111.13714}{arXiv:2111.13714}.

\bibitem{wcrfh25}
Wittmann~W. K., Castro E.R., Roditi I., Foerster A., Hirsch J.G., Subtle
  nuances between quantum and classical regimes,
  \href{https://doi.org/10.1063/5.0237598}{\textit{Chaos}} \textbf{35} (2025),
  043108, 9~pages, \href{http://arxiv.org/abs/2411.07373}{arXiv:2411.07373}.

\bibitem{wyblf23}
Wittmann~W. K., Ymai L.H., Barros B.H.C., Links J., Foerster A., Controlling
  entanglement in a triple-well system of dipolar atoms,
  \href{https://doi.org/10.1103/physreva.108.033313}{\textit{Phys. Rev.~A}}
  \textbf{108} (2023), 033313, 11~pages,
  \href{http://arxiv.org/abs/2305.09754}{arXiv:2305.09754}.

\bibitem{wytlf18}
Wittmann~W. K., Ymai L.H., Foerster A., Tonel A.P., J. L., Control of tunneling
  in an atomtronic switching device,
  \href{https://doi.org/10.1038/s42005-018-0089-1}{\textit{Commun. Phys.}}
  \textbf{1} (2018), 91, 10~pages,
  \href{http://arxiv.org/abs/1710.05831}{arXiv:1710.05831}.

\bibitem{ytfl17}
Ymai L.H., Tonel A.P., Foerster A., Links J., Quantum integrable multi-well
  tunneling models,
  \href{https://doi.org/10.1088/1751-8121/aa7227}{\textit{J.~Phys.~A}}
  \textbf{50} (2017), 264001, 13~pages,
  \href{http://arxiv.org/abs/1606.00816}{arXiv:1606.00816}.

\bibitem{zqwcc24}
Zheng M., Qiao Y., Wang Y., Cao J., Chen S., Exact solution of the
  {B}ose--{H}ubbard model with unidirectional hopping,
  \href{https://doi.org/10.1103/physrevlett.132.086502}{\textit{Phys. Rev.
  Lett.}} \textbf{132} (2024), 086502, 7~pages,
  \href{http://arxiv.org/abs/2305.00439}{arXiv:2305.00439}.

\end{thebibliography}
\end{document}